# Intent-Aware Permission Architecture: A Model for Rethinking Informed Consent for Android Apps

Md Rashedur Rahman[1], Elizabeth Miller[1], Moinul Hossain[1] and Aisha Ali-Gombe[1]
[1]*Department of Computer Information Sciences, Towson University, Towson, Maryland, United States of America*
{mrahma4, emille51}@students.towson.edu, {mhossain, aaligombe}@towson.edu

Abstract: As data privacy continues to be a crucial human-right concern as recognized by the UN, regulatory agencies have demanded developers obtain user permission before accessing user-sensitive data. Mainly through the use of privacy policies statements, developers fulfill their legal requirements to keep users abreast of the requests for their data. In addition, platforms such as Android enforces explicit permission request using the permission model. Nonetheless, recent research has shown that service providers hardly make full disclosure when requesting data in these statements. Neither is the current permission model designed to provide adequate informed consent. Often users have no clear understanding of the reason and scope of usage of the data request. This paper proposes an unambiguous, informed consent process that provides developers with a standardized method for declaring Intent. Our proposed Intent-aware permission architecture extends the current Android permission model with a precise mechanism for full disclosure of purpose and scope limitation. The design of which is based on an ontology study of data requests purposes. The overarching objective of this model is to ensure end-users are adequately informed before making decisions on their data. Additionally, this model has the potential to improve trust between end-users and developers.

## 1 INTRODUCTION

As of 2021, there are 2.8 billion active Android device users, representing 75 percent of the total global mobile users. From social media to online purchases, financial services, banking, music, and entertainment, mobile applications leverage the sophistication of smart devices' sensors coupled with their increasing processing power and mobility to provide the needed functionalities to end-users. A good number of these apps, however, utilize the design of smart devices to store and process valuable and sensitive concrete and behavioral data that can potentially be used to abuse and infringe on user privacy.

By design, Android is built with a permission model that ensures requests to access specific user and device-sensitive data are explicitly granted by the device user at runtime. Although improved over the last few years from the its original designed, the current 1-time fine-grained dynamic permission model is still yet to address two salient user privacy issues related to Intent: 1) purpose limitation and 2) scope of usage or data transfer. It is a well-documented concern in behavioral and usability research that most device users have no clear understanding of why their data was requested by an app. In addition to Android's own permission model, additional solutions were also proposed in the literature to check apps' accesses in exploiting user data (Brutschy et al., 2015; Liu et al., 2016b; Beresford et al., 2011; Zhou et al., 2011; Nauman et al., 2010; Jeon et al., 2012; Bugiel et al., 2013). However, much like the current Android permission model, these solutions have not covered the declaration of the developer's Intent mainly because there were no laws in place that protect the users' rights to be adequately informed. Nevertheless, with the recent expansion of privacy regulations worldwide, particularly the European General Data Protection Regulation (GDPR) and the California Consumer Privacy Act (CCPA), the request for user-sensitive data needs to go through a mandatory and explicit declaration of Intent guided by purpose limitation.

The sections of these current privacy laws related to consent mandated that data request must be made clear and unambiguous. In addition, the regulations further state that the processing of user-sensitive data is limited to the "legitimate purposes" that were "explicitly specified" to the data subject, thus further emphasizing the need for service providers to place and make purpose limitation explicit during consent agreement. Although the use of privacy policy statements to notify users about data requests has been adopted by many Android providers for some time now, however, some of these developers often hide behind complex legal jargon to make ambiguous and, most a times, vague declaration of consent. As documented in the works of (Gluck et al., 2016; Reidenberg et al., 2015), these legal documents are usually hard to parse and understand by average users and even experts.

To address this privacy concern, we argue that a technical implementation that provide a standardize mechanism for developers to explicitly and unambiguously declare Intent programmatically is required. Thus, this research presents a redesign of the Android access permission model to include a fine-grained classification of purposes and mapping purpose-to-scope to create Intent labels, which developers must explicitly declare. Each Intent label defines the purpose for data request and whether or not the data will be used only on the device or transferred. Our proposed Intent-aware permission model extends the current model with a mechanism for developers to comply with the privacy law requirements, particularly the user's right to be informed by clearly defining the purpose and scope limitations at the point of data request. Compared to most exiting works that target permission reduction or privacy enforcement, to the best of our knowledge, this is the first research that addresses the issue of Intent disclosure for mobile apps informed consent. In summary, this paper makes the following contributions:

**1.** An in-depth ontology study of privacy policy statements that examines the terms and language semantics used by experts to declare purpose for data request. This study creates Intent labels which are aggregated into distinctive categories to cover different usage for diverse data groups.

**2.** A new Intent-aware permission model that leverages the developed Intent labels above to create a standardized mechanism for developers to easily comply with the requirement of the law (users' rights to be clearly and unambiguously informed) while imposing very minimal changes to the Android APIs.

**3.** Our model also provides an interface for users to review the Intents (purpose and scope limitations) during permission request, thus equipping them with adequate knowledge and authority to make an informed consent.

The rest of the article is organized as follows: In Section 2, we provide the detailed background of the Android permission model, including its revisions and drawbacks and the motivation for the need for full disclosure and informed consent. Section 3 describes the design and implementation of our new permission model. Section 4 provides a detailed evaluation, discussion, and limitation of the proposed permission model. Section 5 presents the literature review and Section 6 concludes the paper.

## 2 BACKGROUND

### 2.1 Android Permission Model

On Android, user and system applications such as Twitter and Google Maps execute in their sandboxes and maintain their instance of the runtime environment. The kernel layer lies at the bottom of the architecture and manages system resources through the file system, memory and process management, resource allocation, and synchronization. Given the sensitive nature of data and resources hosted on our mobile devices, Android has categorized the different data and resources groups into low-risk and high-risk resources. Generally, any upper-level applications' requests for data and resources are made through the Android permission model which which sits in the middle layer, a bridge between device resources in Kernel space and Application programs. Over the last decade, this permission model has gone through intense scrutiny and modification by the research community and the industry. It has morphed from a flawed all or none-model to a more fine-grained 1-time permission system. Nonetheless, while this system provides program tools for requesting user consent on the data, crucial information about the consent, such as why the data is requested and the scope within which the data will be used, are often ignored. This valuable information on the purpose and scope of the request is excluded entirely in the Permission model, thus leaving the entire consent process without a key crucial component - full disclosure. Much like in medical practice, to explicitly and adequately seek consent on user data, a provider must disclose the data group, access type, purpose, and scope of the data request, however in the current Android Permission model, only the data group is disclosed.

### 2.2 Privacy Policy Statements

The General Data Protection Regulation (GDPR) (GDPR, 2021) of the EU and The California Consumer Privacy Act (CCPA) (CCPA, 2021) of the state of California seeks to address this lack of proper consent and other privacy issues by enacting data protection laws and regulations with specific sections dedicated to the definition of consent (Das et al., 2018). Other bodies such as Brazilian Data Protection Law (LGPD), Personal Information Protection Commission, Japan have also introduced new regulations to protect the privacy of the user while maintaining a fair and cooperative sharing of information with the tech service providers. The commonalities among all these regulations worldwide lie in the requirement of a user's informed, unambiguous, and explicit consent and purpose limitation while sharing certain privacy-sensitive information. To adhere to the requirements of these laws, most services, both online and mobile providers, have adopted the use of privacy pol-icy statements to seek user consent and declare purpose(s) for data requests. However, related work in privacy research has shown that, in an attempt to cover the needed legal concerns by service providers, privacy policies are often drafted as complex, long, and vague statements, thus making it harder for average users to understand (Cate, 2010; Jensen and Potts, 2004; Gluck et al., 2016). This practice is even more complicated on mobile and wearable devices where unclear and lengthy privacy statements are presented to users often one time to seek consent and declare purpose (Gluck et al., 2016). This combination of length and ambiguity in privacy policy statements have rendered the highly worked-out privacy regulations ineffective in protecting average users.

# 3 PROPOSED ANDROID PERMISSION MODEL

As discussed in Section 2, the ambiguity in the consent process poses a significant threat to user privacy. Hence, this paper seeks to address this threat with a simplified and standardized informed consent process, which will require the developers to clearly and explicitly declare the purpose and scope for data requests. Our solution is geared toward improving general user privacy and trust between end-users and developers as well as will be helping developers to comply with the consent section of the new privacy regulations. The design of our proposed approach is threefold - (1) the creation of comprehensive Intent labels that leverages ontology study of data request purposes, (2) the integration of the new model into the Android framework, and (3) Permission UI redesign to include Intent description

## 3.1 Intent Labels

The Intent in the context of this research is a mapping of purpose-to-scope limitation. Specifically, purpose limitation describes the developer's clear intention of why specific data is requested and the associated event or purpose for its usage. In contrast, scope limitation defines the scope within which the developer will use the data, i.e., whether the data will be used on the user device or out of the device. The logical mapping between these two variables describes the developer's Intent. Android being a heterogeneous platform, is associated with multiple different hardware manufacturers and millions of App developers. As such, a simple descriptive approach to defining Intent as adopted by Apple (Hutchinson, ) will lead to the same complex issues we've identified for privacy policy statements, i.e., lengthy, ambiguous, and vague description of purpose and scope. Hence, much like the aggregation of permission, Intent for data requests must be aggregated into distinctive categories that cover different usage for diverse data groups. Therefore, we must define fine-grain ontology of purposes for each data group and then map them to the scope of the request. However, since there is no known ontology for purpose limitation terminologies, we must rely on extensive language analysis of Android apps' privacy policy statements.

### 3.1.1 Ontology Generation

Our goal is to acquire the terms and language used by experts and policy statement writers to declare purpose for data request. Our ontology of purpose workflow is divided into 5 distinct steps: i) Language Extraction, ii) Data Grouping, iii) Aligning Data Cluster to Permission, iv) Purpose Simplification, and v) Purpose Synonymization.

**Phase 1: Language Extraction.** The objective here is to extract language used by Android app developers for purpose declaration into a corpus database. We begin this process with the selection of privacy policy statements for apps that fall into the following categories - i) popular apps (currently covered 80 apps and games) by the Big Techs that maintain proper compliance with the privacy regulations and are expected to share extensive reasons for accessing user data, ii) representative apps from selected categories such as social media, finance, travel and local, shopping etc. This criterion ensures the accuracy of language semantics whilst maintaining diversity.

For each privacy policy statement in our list, we leverage a heuristic approach inspired by the work of Breaux et al. (Breaux and Schaub, 2014) for language decomposition. This approach utilizes a data-verb-purpose mapping to identify the relationship between object, action verb, and a purpose for the data request. With this approach, the first step is to manually deconstruct a privacy policy document into chunks of statements based on an action verb that we termed **Requesting verbs**. Statements containing at least one Requesting verb will be added to a candidate list. The survey of existing literature indicates that the most commonly used action verbs for requesting permission on user data are **collect, access, transfer, share, retain, use** (Breaux and Schaub, 2014; Breaux et al., 2014). Each statement in our candidate list is then further decomposed into three components: the object representing the requested data; the verb, which denotes the type of request; and the event, which shows the purpose for which the data will be used. Next, we extract the purpose description portion of this relation and its corresponding data and add them to the language database. In our study of 80 apps, we extracted more than 1000 statements, which corresponds to 1000 data units and purpose description statements (which are either single words, simple phrases, or complex expressions). Figure 1 illustrates the language decomposition process for a single statement from the TikTok's privacy policy statement.

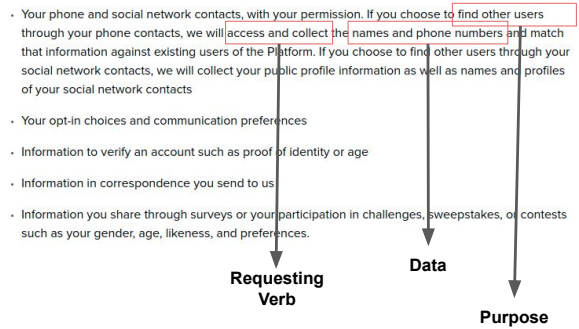

Figure 1: An illustration of Language Extraction of TikTok App's Privacy Policy Statement

**Phase 2: Data Grouping.** Based on the collected data in the corpus database, we group similar data types into clusters. For this purpose, we applied a semantic filtering approach (Doan et al., 2012) to identify and cluster the different expressions of logically related data in our corpus database. For example, in Figure 1, *name* and *phone number* can be simplified as *contact information*. In another example, *Location data* and *Navigation data* have the same semantic meaning in the context of data requests, hence can be logically grouped into a single unit. We created a separate table in the corpus database to store different data groups and their corresponding purpose descriptions. Additional metadata for the name of the app and its category is also included.

**Phase 3: Aligning Data Cluster to Permission.** Next, we establish a relationship between the data clusters, their purpose descriptions, and the standard Android permission labels. This process is crucial to implement the Intent-aware consent process. The mapping allows developers to easily select an Intent label for a particular permission during app development and for users to review the Intent during permission approval at runtime. as shown in Figure 1, the permission needed to access contacts or address book is the *READ_CONTACTS* permission. In another example, user data that belongs to the location cluster (location data, navigation data, etc.) are mapped to the *ACCESS_FINE_LOCATION* and *ACCESS_NETWORK_STATE* permissions. The data grouping table from phase 1 is modified to include the additional column for permissions.

**Phase 4: Purpose Simplification.** Our review of the corpus database showed that the purpose description is often presented either as a single word, a phrase, and in most cases, as a complex expression. In this phase, the goal is to simplify the language and streamline each purpose description. We leveraged the manual grammar reduction technique (Santaholma, 2007) to simplify all complex purpose descriptions. For example, according to one privacy policy, "*we require your data to run or tailor our services, including with more relevant information such as local trends, articles, advertisements, and suggestions for people to follow.*" This purpose is reduced to *personalizing content depending on localization*. In Figure 1, the Tiktok developers declared that- "*we will access and collect the names and phone numbers and match that information against existing users in the platform,*" can be simplified to - *discovering other users on the platform*.

**Phase 5: Purpose Synonymization.** The purpose description language developed in Phase 4 is further simplified by combining purposes that have the same meaning or are closely related. The objective of this step is to reduce information redundancy and generate much simpler purpose statement that conveys clear and succinct intent. In this step, we aim for a one or two-word description for each purpose description. For example, the reduced grammars for *Content personalization*, *personalizing content depending on localization*, *Preferred resources*, and so on are all streamlined into *CONTENT_PERSONALIZATION*. Similarly, the reduced form of purpose taken from Tiktok statement in Figure 1 is reduced to *USER_CONNECT*.

### 3.1.2 Scope Limitation

Scope Limitation is the second attribute of the Intent label for data requests that specify the scope of usage. In this phase, we manually explore the corpus of our privacy policy statement to determine the scope associated with different data groups and Intents from the developer's perspective. Unfortunately, there is a lack of proper disclosure as to whether the collected data will remain on the device or will be transferred even for the categories of apps we chose for analysis. In very few instances, the scope may be implicitly inferred from the requesting verb. For example, where the requesting verb is "transfer", then the scope will assume a definite OFF_DEVICE status. Other ambiguous requesting verbs such as *access*, *use*, *collect*, etc., do not provide adequate clarity as to whether or not the data will be used on the device or otherwise. Thus, we use a combination of heuristic and logical reasoning to identify if a purpose can be achieved strictly ON_DEVICE or OFF_DEVICE. For instance, in the Tiktok statement in Figure 1, the scope is not directly mentioned. However, based on the language analysis, it can be assumed that the contact data may be used both on and off the device; hence, the Scope Limitation considered as OFF_DEVICE.

Using the derived ontology of purpose terms, their data group mapping and the classification of scope limitations above, we then create a strict mapping of purpose-to-scope to form Intent labels. Table 1, illustrate some example of our derived *Intent Labels*.

## 3.2 Framework Modification

The next task in our proposed permission model after the aggregation and creation of Intent labels (purpose-to-scope mapping) is to modify the Android framework so that a developer can leverage our Intent labels to explicitly declare purpose and scope when requesting permission to sensitive user data. The Android framework is a set of APIs that allows developers to create apps and interact with the user, device and access user data and resources. In the current framework, access to specific user-sensitive data is made through specialized APIs such as the *requestPermission()* function. These function calls currently take two parameters—the permission and the request mode.

In our new permission model, we modify the framework such that additional data structures that hold the Intent labels are defined. Next, we modify the *requestPermission* functions to include three additional parameters for the Intent labels (purpose and scope) and user-specific purpose descriptions. We then create a program mapping between permissions and Intent labels such that when a developer chooses specific permission, an associated list of approved purposes for that specific permission will be populated. These changes are introduced primarily in the internal framework code with minimal changes to the Android developer's SDK. During app development, alongside the permission, the developer can choose from the available Intent label the purpose and scope limitations of their data requests.

## 3.3 Permission UI Redesign

The primary objective of the new permission model is to ensure users are better informed about why their data is requested and the scope of data usage. Thus, the final task in the design of this model is to modify the permission check UI such that additional information for the Intent label is displayed to the user at runtime. In the current model, permission request API calls trigger a dialogue box for a user to accept or deny permission. However, the user is not provided with the purpose and scope of the data request. Thus, in our proposed model, we added another dialogue box to render the Intent label (purpose-to-scope limitation) along with the Permission Name. Furthermore, a developer-specific purpose description or link to complete privacy statement will also be shown to the user. We will maintain the two buttons for accepting or denying the permission. This would give a comprehensive depiction of the overall purpose of the request to the user. The user gets to allow the request based on their satisfaction with the developer's chosen Intent. In general, these changes to the permission check dialogue would aid users in making an informed decision while accepting or denying the permission.

## 3.4 Implementation

We implemented a prototype of our new permission model on version 10 of the Android Open Source Project (AOSP) (Android, 2021). Firstly, we added two new Enums called - *PermissionReasonEnum* and *DataScopeEnum* to address the purpose and scope limitations respectively. In the current implementation, *PermissionReasonEnum* and *DataScopeEnum* are formulated based on the use cases of the following privacy sensitive resources: *Location Data*, *SMS*, *Camera*, *Contact*, *Device ID*, and *Call Logs*. We would be extending this implementation further in the future to address every privacy sensitive data types in the Android System. Next, we create a new class called *PermissionWithReason* with four member variables—String *permissionName*, PermissionReasonEnum *permissionReasonTitle*, String *permissionReasonDescription*, and DataScopeEnum *dataScope*. Using case-switch conditional statements, the implementation of the *PermissionWithReason* forces the desired mapping for the Intent labels i.e., map certain purpose to restricted scope.

In our current implementation, we utilized method overloading of the requestPermission() function (which is a scenario of having more than one method with the same name, but different signatures) to address the additional definition of Intent and backward compatibility to the existing framework. As mentioned earlier, in the current framework, the *requestPermission* function takes two parameters—a string array of permission names and code. We then overload this method with an additional implementation that takes an ArrayList of *PermissionWithReason* class as an additional parameter. To properly declare Intent, developers are required to use this new function during app development. They will first need to create an instance(s) of the *PermissionWithReason* with their permission name, and then choose the appropriate enum for purpose and scope and then provide a description of the purpose in string form. Multiple instances of this class can be created for different permissions and then passed to the *requestPermission* function in an ArrayList.

# 4 EVALUATION

## 4.1 Evaluation

To test the effectiveness and usability of our proposed permission model in terms of functionality and compatibility, we conducted two types of evaluations—

Table 1: Intent Label to permission mapping.

| Data Type | Permission | Intent | Scope |
|---|---|---|---|
| Fingerprint | USE_FINGERPRINT | AUTHENTICATION | ON_DEVICE |
| | | SECURITY | ON_DEVICE |
| Location | ACCESS_FINE_LOCATION | ADVERTISEMENT | OFF_DEVICE |
| | | APP_EXPERIENCE | OFF_DEVICE |
| | | APP_SERVICE | OFF_DEVICE |
| | | CONTENT_PERSONALIZATION | OFF_DEVICE |
| | | DIAGNOSTICS | OFF_DEVICE |
| | | ENHANCED_SERVICE | OFF_DEVICE |
| | | FRAUD_DETECTION | OFF_DEVICE |
| | | PERSONALIZED_OFFERS | OFF_DEVICE |
| | | TRACKING | OFF_DEVICE |

functional and performance testings. Our evaluation method seeks to answer the following questions:

- How easy is it for developers to adopt the proposed permission model for new app development and backward compatibility with existing apps built with an old framework?

- What is the performance impact of the new model on a device after the framework modification?

### 4.1.1 Functional Testing

The functional testing experiment provides a clear description of how our new permission model can be built into the exiting Android framework and deployed onto a device. Additionally, this test answers how easy it is for developers to adopt our framework and its support for backward compatibility.

We implemented our proposed framework on Android 10 using the Android Open Source Project. After adding all the required data types, classes, and functions to the project, we build and compiled a new system image containing the new framework code. Additionally, an SDK containing the new *requestPermission* method call is also generated. This new system image is then flashed into an already created Android Virtual Device (AVD), and the path to the new SDK is added to the Android studio. Due to technical and legal constraints in ROM modification for new Android devices (LionGuest-Studios, 2021; DaveTheTytnIIGuy, ), we built and test our prototype only on an AVD. Nonetheless, we are confident that our model can easily be adopted and deployed by Android and/or any Hardware manufacturer on a real device without any challenges.

**New App Development.** To test the ease of use for developers, we develop a Test app called "SampleGPSTesting". "SampleGPSTesting" is a location detection app that shows current *Latitude* and *Longitude* value based on the user's current location. By design, this app require LOCATION data and INTERNET access to accomplish its app functionality, and hence it request for the *android.permission.ACCESS_FINE_LOCATION* and *android.permission.-INTERNET* permissions from the user in the first activity. Thus, in developing this program, we called the *requestPermission* function with four parameters supplying the *context*, *list_of_permissions*, *requestCode*, List of *permissionWithReasons*. This app is compiled and executed in the emulator. Figure 2 is a screenshot of the permission UI dialogue box showing purpose limitation (NAVIGATION), developer-specific purpose description (To access...), and scope limitation (ON_DEVICE) of the location data requested by the SampleGPSTesting.

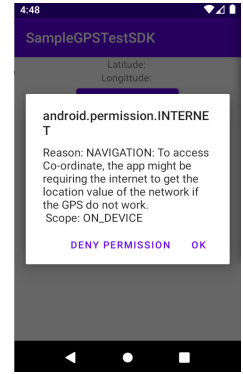

Figure 2: A screenshot of the Permission UI showing the Intent labels and description for the SampleGPSTesting app.

**Modification of Exiting App.** In this experiment, we aim to examine how easy is it for developers to modify existing apps to incorporate the additional information for purpose and scope during permission requests. We use a case study-based approach where we modify an existing open-source Android application—*Phonograph* (AbouZeid, 2020)—to directly use our proposed framework. We downloaded the source code from Github and examined the permission list built into its manifest file. This app is designed as a music player and has the following permission list WAKE_LOCK, READ_EXTERNAL_STORAGE etc. Based on our current implementation of the new permission model, only the READ_EXTERNAL_STORAGE permission accesses sensitive data, and thus we mapped this permission to its API call and then manually trace the API call to identify why access to read the device storage is requested. Our evaluation shows that the Phonograph app must read songs from an external storage unit on the device before playing it. Thus, with this deducted reasoning, we modified the app's permission requesting functions to use our new requestPermission function with PLAY_MUSIC purpose and an ON_DEVICE scope. After compiling and generation the apk, we installed the app on our AVD that runs the modified version of our framework. As shown in Figure 3, the modified app displayed to the user the requested permission, its purpose limitation, description, and scope limitation. This additional information will help users of the Phonograph app in making informed decisions whether to allow or deny access to read their external storage.

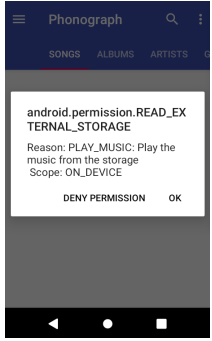

Figure 3: The screenshot of the Permission UI showing the Intent labels and description for *Phonograph* after it was modified to use the proposed model.

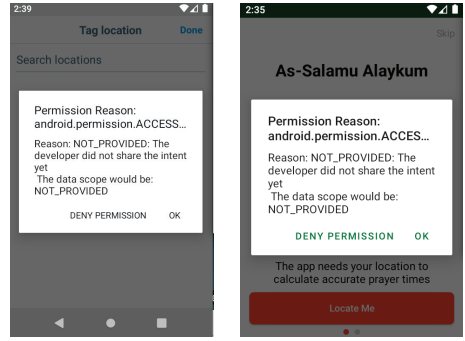

(a) **Twitter**   (b) **Muslim Pro**

Figure 4: The screenshots of the Permission UI showing the Intent labels and description for currently exiting applications that were compiled with old SDKs.

**Backward Compatibility.** In the second case study, we aim to address backward compatibility with existing apps that use the old framework. In real-world application scenarios, if Android adopts this permission model, we can expect all new apps to immediately follow our proposed new model. However, it will take some time for exiting apps to adapt and modify, and as such old framework may continue to exist before being permanently phased out. Thus, our framework is designed to support backward compatibility and ensure that our new framework does not break currently exiting applications that were compiled with old SDKs. To analyze this requirement, we installed some existing apps from Google PlayStore in our framework-modified AVD. Our results suggest none of the apps break. All the apps functioned as expected on the new image. Figure 4 showed the screenshots for two apps: Twitter (Twitter, 2021) and Muslim Pro (MuslimPro, 2021). Because both apps were not compiled with the new framework, the purpose and scope have defaulted to the NOT_PROVIDED.

These case studies conclude that developers can easily adapt our proposed Framework to create new apps and update their existing apps. Additionally, the last test shows our model is backward compatible with unmodified apps.

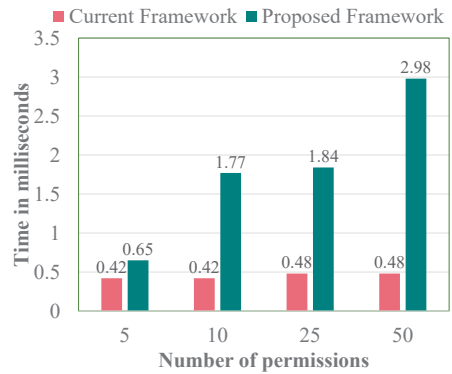

Figure 5: Performance chart.

### 4.1.2 Performance Evaluation

Our framework modification and UI redesign phases have the potential of imposing some performance overheads particularly in terms of app's start time. In this evaluation we seek to determine whether there would be any significant delay in the activity start time, which can degrade the user's satisfaction with the system. We developed a demo app containing 5, 10, 25, and 50 permissions with corresponding purposes and scopes to conduct this test. We change the number of permissions the app asks for each execution and record the activity start time for both the unmodified framework and our proposed Android Framework. The Intent parameters are only provided when executing the apps in the proposed framework. All the performance tests were conducted on an AVD with 16 GB RAM.

Figure 5 shows the comparison between the app's initial start time when using the existing framework versus the proposed framework. As expected, there is a slight increase in the app's initial start time, which also increases linearly with an increase in the number of permissions. An analysis of this performance penalty suggests that the additional loop added to generate the AlertBox and the conditional mapping of purpose-to-scope limitation are responsible for in-creasing the app's initial start time. Given that the average requested permission for Android apps as documented by Pew Research is 5, this performance penalty will be very negligible for most apps.

## 4.2 Discussion and Limitation

Our evaluation results showed the proposed framework can significantly improve user data privacy on Android applications by enforcing current privacy laws. In addition, we have shown that our implementation is easy to build and deploy by device manufactures and can easily be adopted by developers. Furthermore, our backward compatibility test showed that exiting apps can still work with the new framework. While our implementation has addressed some critical challenges, it has some important limitations that we hope to address in future work.

- Blind Trust: Currently, the existing system blindly trusts that the developer will only be using the required data for the declared purpose, and there is no mechanism to check if the developer choose to use the data otherwise. Thus, in our future work, we hope to develop a more trustful permission architecture using memory analysis backward propagation to ensure that data requested for a particular purpose and on-device scope are enforced.

- Multi-Purpose Request: Our current implementation is limited to a single purpose. We plan to implement a multi-purpose scope to improve the usability of the proposed permission model.

- Usability Testing: Our evaluation is focused entirely on the functional architecture of our proposed Permission model. Given that this is a user privacy-preserving technique, the usability of the system between users from various groups and backgrounds would be investigated as part of our future work. In addition, the usability tests on developers will also be conducted to understand their adaptability and comfort in the proposed model. These testings will help us create a more usable dialogue prompt, streamlined purpose categorization, and Intent labels.

# 5 Related Work

## 5.1 Low-Level Modification

Several studies have explored the introduction of fine controls over permissions at the framework, the middle layer, or the operating system-level using different strategies (Beresford et al., 2011; Zhou et al., 2011; Brutschy et al., 2015; Nauman et al., 2010; Jeon et al., 2012; Liu et al., 2016b; Caputo et al., 2021). Proposed techniques such as Mockdroid (Beresford et al., 2011), and TISSA (Zhou et al., 2011) facilitate users to control access whenever access to the resource is attempted by providing dummy information instead of the real one. This strategy allows most applications to continue execution at the expense of potential usefulness. Apex (Nauman et al., 2010) propose a mechanism for users to carefully grant permissions to apps based on runtime contextual information, such as the device location or how frequently a resource has been previously used. Other solutions such as (Hornyack et al., 2011; Shinde and Sambare, 2015; Liu et al., 2016a; Rashidi et al., 2016; Chen et al., 2017; Raval et al., 2019) proposed permission configuration interface that will allow users to control the activities of third-party apps. FlaskDroid (Bugiel et al., 2013) on the other hand, provides compulsory access control on Android's middleware and kernel layers to avert unwanted information disclosure. Some recent solutions, such as (Caputo et al., 2021) focused on anonymization of data to preserve the privacy of the user. In (Qu et al., 2020), authors introduce an automatic permission management model, which collectively considers the risk of the permission, functionalities of the app, and user's privacy requirement. However, none of these solutions helps users understand the Intent of the developer. Furthermore, there is no policy integrated into the fabric of Android OS to reveal the Intent behind each permission, making it more difficult for users to make informed decisions.

## 5.2 Application-Level Modification

AppIntent (Yang et al., 2013) proposes an application-level solution that aims to control information dissemination by identifying if the user intends the transmission. Though this solution is much related to our proposed method, it is a reactive approach that does not consider introducing the developer's Intent for the data request in the Android permission process. In contrast, our work takes a proactive approach that provides end-users with an explicit declaration of the developer's Intent upfront during permission requests. Other app-level fine-grain permission or permission reduction solutions are (Jeon et al., 2012; Backes et al., 2012; Backes et al., 2013; Davis and Chen, 2013; Zhang and Yin, 2014; Brutschy et al., 2015; Ali-Gombe et al., 2016). Dr. Android (Jeon et al., 2012) retrofits apps to allow for the specification of fine-grained variants of Android permissions by accumulating existing permission into a taxonomy, each of which admits some common strategies for deriving sub-permissions. In (Do et al., 2014; Chen et al., 2019), the authors introduce a permission removal approach to mitigate privacy leaks in Android. Appguard (Backes et al., 2012; Backes et al., 2013) and RetroSkeleton (Davis and Chen, 2013) are both an app-agnostic flexible and dynamic policy re-writing solutions designed towards enhancing security and privacy in untrusted Android applications. Capper (Zhang and Yin, 2014) is an app-level taint tracking system that examines sensitive data flow from source to sink. SHAMDROID (Brutschy et al., 2015) proposes a permission slicing, and reduction solution aims at degrading apps functionality in the resource constraint systems like Android. priVy(Ali-Gombe et al., 2016) is a portable, fine-grained access control solution designed to achieve permission reduction in the Android native content providers. While these solutions are geared towards permission reduction and enhancing user privacy, unlike our proposed solution, these solutions do not explicitly state the developer's purpose and scope for requesting the user data, thus hindering users' ability to make an informed decision. Other related work such as Flowverine provides an app developer with the tools to build a privacy-aware and secured app (Gomes et al., 2020). Although this mechanism allows the system to track a certain app's data flow, it does not give the user control over their data. Configurable User Privacy Approach (CUPA) (Alkindi et al., 2020), proposes a configuration mechanism for users to have complete control over the data flow of the app and can intercept if necessary. This solution, however, lies in the lack of technical knowledge of the user to define the required policies. In contrast, our solution builds an aggregated labels for Intent based on a extensive ontology study, which are then used for defining purpose and scope limitations. The combination of the proposed Intent specification, scope, and description makes it easier for users to decide whether to allow or deny the request for their privacy-sensitive data at runtime.

## 6 CONCLUSION

User data privacy is a fundamental human right that tech companies and service providers, particularly mobile providers, often violate and abuse. In this paper, we extended the current Android permission model to address this issue. Our model ensures users can make an informed decision with adequate information before approving data access on Android devices. The design of our model includes a manual aggregation of request purposes which is mapped to scope limitation to form Intent labels. These Intent labels are then built into the new permission model for easy categorization of purpose-to-scope limitation. Our evaluation showed the new model could aid apps to comply with the requirements of the current regulations while incurring very minimal performance overhead. In addition, the minimal changes required by the developer make the adoption of this model easy while still providing backward compatibility for non-complying apps.

## ACKNOWLEDGEMENTS

This work is supported by the National Science Foundation (NSF) Grant Number 1850054 and Towson University OSPR Pilot Research Seed Grant.

## REFERENCES

AbouZeid, K. (2020). Phonograph - A material designed local music player for Android. https://github.com/kabouzeid/Phonograph. (Accessed on June 15, 2021).

Ali-Gombe, A., Richard III, G. G., Ahmed, I., and Roussev, V. (2016). Don't touch that column: Portable, fine-grained access control for android's native content providers. In *Proceedings of the 9th ACM Conference on Security & Privacy in Wireless and Mobile Networks*, pages 79–90.

Alkindi, Z., Sarrab, M., and Alzidi, N. (2020). Cupa: A configurable user privacy approach for android mobile application. In *2020 7th IEEE International Conference on Cyber Security and Cloud Computing (CSCloud)/2020 6th IEEE International Conference on Edge Computing and Scalable Cloud (EdgeCom)*, pages 216–221. IEEE.

Android (2021). Android open source project. https://source.android.com/. (Accessed on June 1, 2021).

Backes, M., Gerling, S., Hammer, C., Maffei, M., and von Styp-Rekowsky, P. (2012). Appguard-real-time policy enforcement for third-party applications.

Backes, M., Gerling, S., Hammer, C., Maffei, M., and von Styp-Rekowsky, P. (2013). Appguard–enforcing user requirements on android apps. In *International Conference on TOOLS and Algorithms for the Construction and Analysis of Systems*, pages 543–548. Springer.

Beresford, A. R., Rice, A., Skehin, N., and Sohan, R. (2011). Mockdroid: trading privacy for application functionality on smartphones. In *Proceedings of the 12th workshop on mobile computing systems and applications*, pages 49–54.

Breaux, T. D., Hibshi, H., and Rao, A. (2014). Eddy, a formal language for specifying and analyzing data flow specifications for conflicting privacy requirements. *Requirements Engineering*, 19(3):281–307.

Breaux, T. D. and Schaub, F. (2014). Scaling requirements extraction to the crowd: Experiments with privacy policies. In *2014 IEEE 22nd International Requirements Engineering Conference (RE)*, pages 163–172. IEEE.

Brutschy, L., Ferrara, P., Tripp, O., and Pistoia, M. (2015). Shamdroid: gracefully degrading functionality in the presence of limited resource access. *ACM SIGPLAN Notices*, 50(10):316–331.

Bugiel, S., Heuser, S., and Sadeghi, A.-R. (2013). Flexible and fine-grained mandatory access control on android for diverse security and privacy policies. In *22nd USENIX Security Symposium*, pages 131–146.

Caputo, D., Pagano, F., Bottino, G., Verderame, L., and Merlo, A. (2021). You can't always get what you want: towards user-controlled privacy on android. *arXiv preprint arXiv:2106.02483*.

Cate, F. H. (2010). The limits of notice and choice. *IEEE Security & Privacy*, 8(2):59–62.

CCPA (2021). California consumer privacy act (ccpa). https://oag.ca.gov/privacy/ccpa. (Accessed on May 29, 2021).

Chen, X., Huang, H., Zhu, S., Li, Q., and Guan, Q. (2017). Sweetdroid: Toward a context-sensitive privacy policy enforcement framework for android os. In *Proceedings of the 2017 on Workshop on Privacy in the Electronic Society*, pages 75–86.

Chen, Y., Zha, M., Zhang, N., Xu, D., Zhao, Q., Feng, X., Yuan, K., Suya, F., Tian, Y., Chen, K., et al. (2019). Demystifying hidden privacy settings in mobile apps. In *Proceedings of IEEE Symposium on Security and Privacy*, pages 570–586. IEEE.

Das, A., Acar, G., Borisov, N., and Pradeep, A. (2018). The web's sixth sense: A study of scripts accessing smartphone sensors. In *Proceedings of the 2018 ACM SIGSAC Conference on Computer and Communications Security*, pages 1515–1532.

DaveTheTytnIIGuy. Is flashing roms legal? well i went straight to the big guys. https://forum.xda-developers.com/t/is-flashing-roms-legal-well-i-went-straight-to-the-big-guys.598449/. (Accessed on June 15, 2021).

Davis, B. and Chen, H. (2013). Retroskeleton: Retrofitting android apps. In *Proceeding of the 11th annual international conference on Mobile systems, applications, and services*, pages 181–192.

Do, Q., Martini, B., and Choo, K.-K. R. (2014). Enhancing user privacy on android mobile devices via permissions removal. In *Proceedings of the 47th Hawaii International Conference on System Sciences*, pages 5070–5079. IEEE.

Doan, S., Ohno-Machado, L., and Collier, N. (2012). Enhancing twitter data analysis with simple semantic filtering: Example in tracking influenza-like illnesses. In *2012 iEEE second international conference on healthcare informatics, imaging and systems biology*, pages 62–71. IEEE.

GDPR (2021). General data protection regulation- gdpr. https://gdpr-info.eu/. (Accessed on May 29, 2021).

Gluck, J., Schaub, F., Friedman, A., Habib, H., Sadeh, N., Cranor, L. F., and Agarwal, Y. (2016). How short is too short? implications of length and framing on the effectiveness of privacy notices. In *Twelfth Symposium on Usable Privacy and Security ({SOUPS} 2016)*, pages 321–340.

Gomes, E., Zavalyshyn, I., Santos, N., Silva, J., and Legay, A. (2020). Flowverine: Leveraging dataflow programming for building privacy-sensitive android applications. In *Proceedings of 19th IEEE International Conference On Trust, Security And Privacy In Computing And Communications (TrustCom)(to appear)*.

Hornyack, P., Han, S., Jung, J., Schechter, S., and Wetherall, D. (2011). These aren't the droids you're looking for: retrofitting android to protect data from imperious applications. In *Proceedings of the 18th ACM conference on Computer and communications security*, pages 639–652.

Hutchinson, A. Apple officially launches new app tracking permissions as part of ios 14.5. https://www.socialmediatoday.com/news/apple-officially-launches-new-app-tracking-permissions-as-part-of-ios-145/599057/. (Accessed on June 15, 2021).

Jensen, C. and Potts, C. (2004). Privacy policies as decision-making tools: an evaluation of online privacy notices. In *Proceedings of the SIGCHI conference on Human Factors in Computing Systems*, pages 471–478.

Jeon, J., Micinski, K. K., Vaughan, J. A., Fogel, A., Reddy, N., Foster, J. S., and Millstein, T. (2012). Dr. android and mr. hide: fine-grained permissions in android applications. In *Proceedings of the second ACM workshop on Security and privacy in smartphones and mobile devices*, pages 3–14.

LionGuest-Studios (2021). Method overloading. https://www.youtube.com/watch?v=ffsyQcSOnFw. (Accessed on June 15, 2021).

Liu, B., Andersen, M. S., Schaub, F., Almuhimedi, H., Zhang, S. A., Sadeh, N., Agarwal, Y., and Acquisti, A. (2016a). Follow my recommendations: A personalized privacy assistant for mobile app permissions. In *Twelfth Symposium on Usable Privacy and Security ({SOUPS} 2016)*, pages 27–41.

Liu, R., Liang, J., Cao, J., Zhang, K., Gao, W., Yang, L., and Yu, R. (2016b). Understanding mobile users' privacy expectations: A recommendation-based method through crowdsourcing. *IEEE Transactions on Services Computing*, 12(2):304–318.

MuslimPro (2021). The most popular muslim app! https://www.muslimpro.com/. (Accessed on June 1, 2021).

Nauman, M., Khan, S., and Zhang, X. (2010). Apex: extending android permission model and enforcement with user-defined runtime constraints. In *Proceedings of the 5th ACM symposium on information, computer and communications security*, pages 328–332.

Qu, Y., Du, S., Li, S., Meng, Y., Zhang, L., and Zhu, H. (2020). Automatic permission optimization framework for privacy enhancement of mobile applications. *IEEE Internet of Things Journal*.

Rashidi, B., Fung, C., and Vu, T. (2016). Android fine-grained permission control system with real-time expert recommendations. *Pervasive and Mobile Computing*, 32:62–77.

Raval, N., Razeen, A., Machanavajjhala, A., Cox, L. P., and Warfield, A. (2019). Permissions plugins as android apps. In *Proceedings of the 17th Annual International Conference on Mobile Systems, Applications, and Services*, pages 180–192.

Reidenberg, J. R., Breaux, T., Cranor, L. F., French, B., Grannis, A., Graves, J. T., Liu, F., McDonald, A., Norton, T. B., and Ramanath, R. (2015). Disagreeable privacy policies: Mismatches between meaning and users' understanding. *Berkeley Tech. LJ*, 30:39.

Santaholma, M. E. (2007). Grammar sharing techniques for rule-based multilingual nlp systems. In *Proceedings of the 16th Nordic Conference of Computational Linguistics (NODALIDA)*.

Shinde, S. S. and Sambare, S. S. (2015). Enhancement on privacy permission management for android apps. In *2015 Global Conference on Communication Technologies (GCCT)*, pages 838–842. IEEE.

Twitter (2021). Twitter. https://twitter.com/. (Accessed on June 1, 2021).

Yang, Z., Yang, M., Zhang, Y., Gu, G., Ning, P., and Wang, X. S. (2013). Appintent: Analyzing sensitive data transmission in android for privacy leakage detection. In *Proceedings of ACM SIGSAC conference on Computer & communications security*, pages 1043–1054.

Zhang, M. and Yin, H. (2014). Efficient, context-aware privacy leakage confinement for android applications without firmware modding. In *Proceedings of the 9th ACM symposium on Information, computer and communications security*, pages 259–270.

Zhou, Y., Zhang, X., Jiang, X., and Freeh, V. W. (2011). Taming information-stealing smartphone applications (on android). In *Proceedings of International conference on Trust and trustworthy computing*, pages 93–107. Springer.